\documentclass[twocolumn,aps,pra,amssymb,showpacs,longbibliography,10pt]{revtex4-1}

\usepackage{amsfonts,amssymb,amsmath,graphicx,color,hyperref}

\newcommand{\br}{{\bf r}}
\newcommand{\vq}{{\bf r}}

\newcommand{\vk}{{\bf k}}
\newcommand{\vsigma}{\boldsymbol{\sigma}}

\definecolor{greenPR}{rgb}{0.00, 0.6, 0.00}
\newcommand{\commentout}[1]{}

\begin{document}

\title{Dirac quantum time mirror}
\author{Phillipp Reck$^1$, Cosimo Gorini$^1$, Arseni Goussev$^2$, Viktor Krueckl$^1$, Mathias Fink$^3$, Klaus Richter$^{1}$}
%\email[]{klaus.richter@ur.de}
\affiliation{$^1$Institut f\"ur Theoretische Physik, Universit\"at Regensburg, 93040 Regensburg, Germany}
\affiliation{$^2$Department of Mathematics, Physics and Electrical Engineering, Northumbria University, 
Newcastle upon Tyne, NE1 8ST, United Kingdom}
\affiliation{$^3$Institut Langevin, ESPCI, CNRS, PSL Research University, 1 rue Jussieu, 75005, Paris, France }
\date{\today}

\begin{abstract}
Both metaphysical and practical considerations related to time inversion have intrigued scientists for generations.
Physicists have strived to devise and implement time-inversion protocols,
in particular different forms of ``time mirrors'' for classical waves.
Here we propose an instantaneous time mirror for {\it quantum} systems,
{\it i.e.} a controlled time discontinuity generating wave function echoes with high fidelities.
This concept exploits coherent particle-hole oscillations in a Dirac spectrum
in order to achieve population reversal, and can be implemented in systems such as (real or artificial) graphene.

\end{abstract}

\maketitle

%%%%%%%%%%%%%%%%%%%%%%%%%%%%%%%% Intro %%%%%%%%%%%%%%%%%%%%%%%%%%%%%%%%%%%%%%%%%%%%%%%%%%%%

\section{Introduction}
\label{sec_intro}

The physicists' fascination with time inversion goes back a long time, 
as testified by the famous 19th-century argument between Loschmidt and Boltzmann concerning the arrow of time \cite{loschmidt1876,boltzmann1877}.
Deep theoretical and metaphysical considerations are not the sole reasons behind it, though.
The pioneering work of Hahn in 1950 \cite{hahn1950}, in which the dynamics of an ensemble of nuclear spins
was successfully time inverted, gave birth to the concept of spin echo,
now central to numerous imaging techniques \cite{EPRbook}. 
A spin echo, at least in its most basic form, can be understood in terms of ``population reversal''
in two-level systems: an ensemble of initially uniformly aligned spins precesses around an applied
magnetic field, progressively losing relative phase coherence; a microwave $\pi$-pulse is then used
to simultaneously flip the spins, making them effectively evolve ``back in time'' regaining (``echoing'') the initially aligned, 
phase coherent configuration.

Another successful approach to time inversion has been developed for classical waves based on time-reversal mirrors implemented with acoustic \cite{fink1992,draeger1997}, elastic \cite{fink1997},
electromagnetic \cite{lerosey2004} and recently water waves \cite{przadka2012,chabchoub2014}.
It relies on the fact that any wave field can be completely determined in a volume by knowing only the field at any enclosing
surface (a spatial boundary). It requires the use of receiver-emitter antennas positioned on the surface that record an incident
wavefront and later rebroadcast a time-inverted copy of the signal. If an initially localized pulse, e.g. a wave emitted from any
source, is left to evolve for a certain time and then in this manner t-inverted on a boundary, it can trace its way back to the
initial source and there refocus or ``echo'' \cite{pastawski2007,calvo2008,calvo2010}. This process is difficult to implement in
optics because of the lack of controllable antennas \cite{mosk2012}, and a solution to create time-reversed waves is to
work with monochromatic light and use three- or four-wave mixing \cite{yariv1978,miller1980}.

A potent alternative to wave field control via spatial boundaries is the manipulation of time boundaries
\cite{moshinsky1952,gerasimov1976,caslav1997,mendonca2002,delcampo2009,goussev2012a,haslinger2013}.   
Specific time reversal protocols for one-dimensional (1D) propagation were proposed \cite{martin2008, martin2008b} and
experimentally realized \cite{ullah2011} in the kicked rotator model of atomic
matter waves (for a narrow range of momenta), 
and proposed for electromagnetic waves \cite{sivan2011a, sivan2011b}, the latter based on time- and space-modulated perturbations of a photonic crystal with linear dispersion.
The latest development in this context is the concept of an instantaneous time mirror, 
which has been verified experimentally \cite{bacot2015} in the field of gravity-capillary waves: A sudden modification of water wave celerity obtained from a vertical acceleration of a bath of water creates a time-reversed wave. 
This time disruption realizes an instantaneous time mirror in the entire space. 
Such a mirror can be viewed as the analogue in time to a standard mirror that acts on space. 

In spite of these successes for classical waves, a long standing challenge remains: are Quantum Time Mirrors (QTMs) feasible 
for spatially extended quantum waves?  
In other words, can one time-invert the motion of a quantum wave propagating in space? 
Notice that unlike spin echoes, which deal with (an ensemble of) discrete two-level systems,
one is here dealing with a continuous degree of freedom describing a wave function extending and evolving coherently in space.
A direct adaption of the aforementioned classical wave strategies appears difficult: On the one hand,
recording and properly re-emitting waves would require to measure and thereby massively change the quantum state;
on the other hand, the intriguing concept of a time mirror for water waves cannot be transferred
to wave functions due to the inherently different structure of the underlying differential equations describing classical and quantum wave propagation. 

\begin{figure*}[t]
   \includegraphics[width=\textwidth]{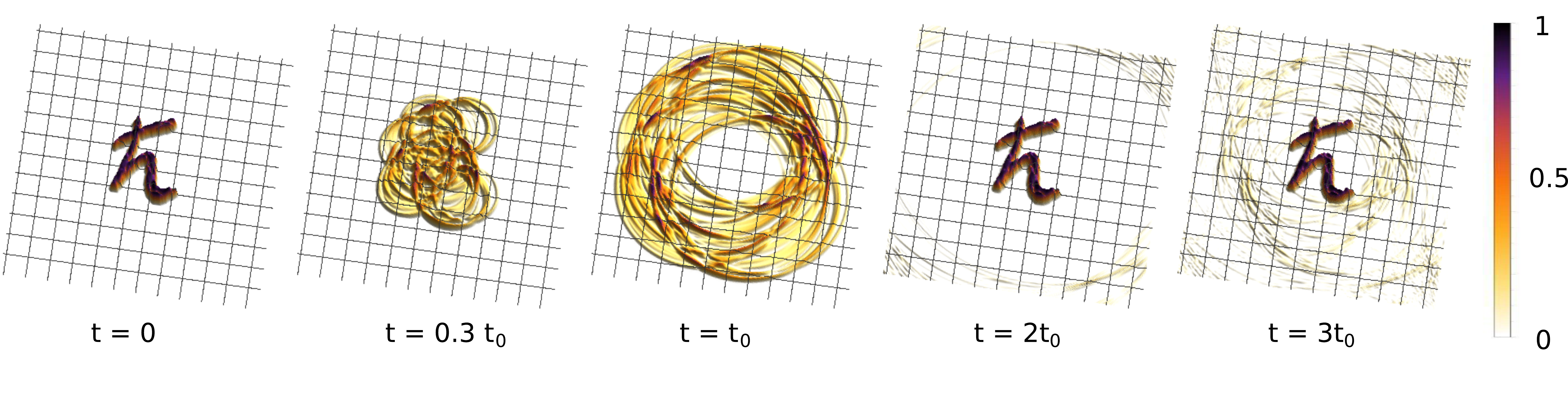}
\caption{Echo in a Dirac quantum system. The absolute value of a $\hbar$-shaped wave packet is shown in real space (a.u.). The intial wave packet (at $t=0$) becomes completely blurred while propagating until $t=t_0$. A fast quantum time reversal pulse at $t=t_0$ leads to a nearly perfect echo at $t=2t_0$. The right panel (at $t=3t_0$) shows a second echo after a further subsequent pulse at $t=2.5 t_0$. This simulation is done without disorder.  The colorscale is the same in all snapshots, normalized to the highest value of the modulus of the initial wave packet.} \label{fig:hbar}
\end{figure*}

Here we show that it is indeed possible to devise high-fidelity (instantaneous) QTMs for the time evolution of wave functions.
We propose a concept of QTMs based on (bosonic or fermionic) Dirac-like systems, such as graphene,
exploiting the ``population reversal'' principle at the heart of spin echoes.  
Indeed, this approach unifies two up to now distinct paradigms,
the $t$-inversion of a spatially extended wave and the generation of a (pseudo)spin echo.
Figure \ref{fig:hbar} gives a taste of its effectiveness: an initially $\hbar$-shaped wave packet
evolves in time progressively losing its profile, until the action of the instantaneous QTM, a short pulse at $t=t_0$,
inverts the propagation and leads to a distinct echo at $t=2t_0$;
the subsequent echo at $t=3t_0$ is due to a further QTM pulse at $t=2.5t_0$
(see movie in \cite{SuppMat}).
Experimentally there are various ways of injecting electronic wave packets -- of simpler shape -- into a system,
notable ones including quantum dots as single-electron sources \cite{bocquillon2013}
or short voltage (``Leviton'') pulses \cite{jullien2014}.

%%%%%%%%%%%%%%%%%%%%%%%%%%%%%%%% Graphene %%%%%%%%%%%%%%%%%%%%%%%%%%%%%%%%%%%%%%%%%%%%%%%%%%%%

\section{Quantum time mirror - basics}
\label{sec_results}

% \section{Quantum Time Mirror for graphene-like systems}
% \label{subsec_graphene}

\begin{figure*}[!tbp]
 \includegraphics[width=0.8\textwidth]{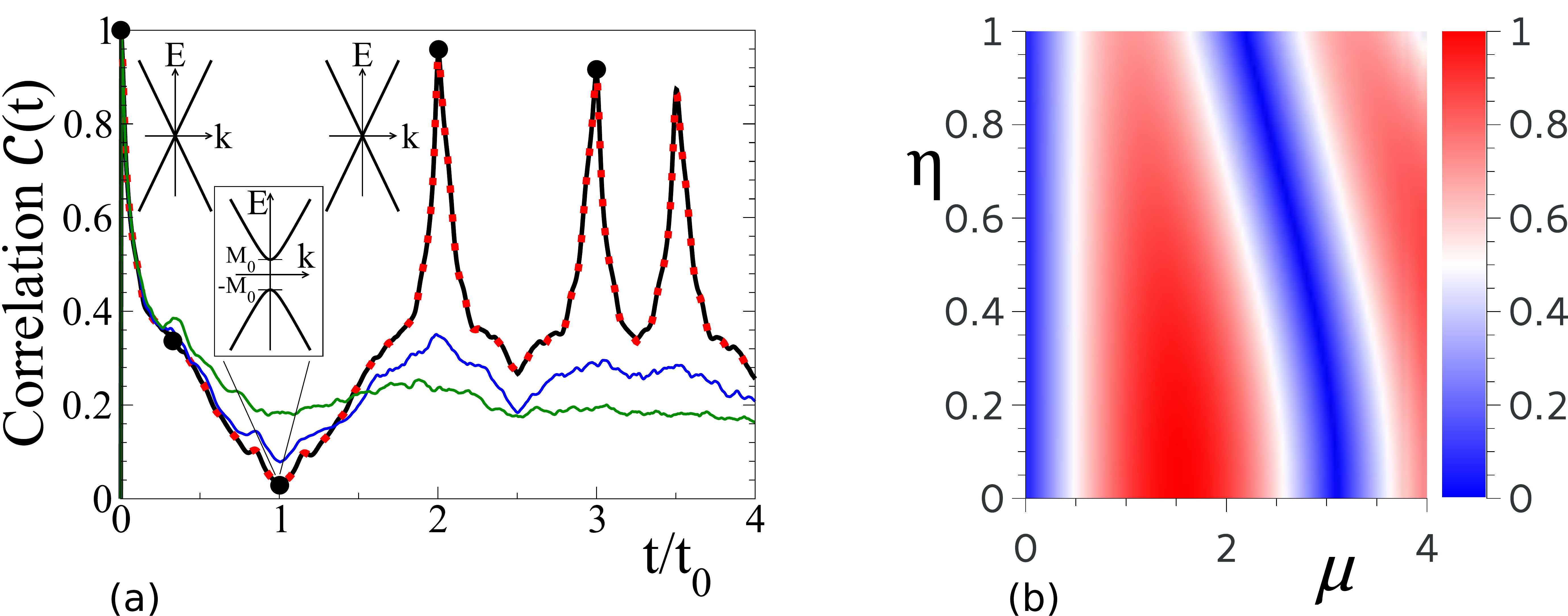}
\caption{Quantitative analysis of the echo strength. (a) Correlation $\mathcal{C}(t)$, Eq.~\eqref{eq:spaceCorr}, obtained from numerical propagation of the $\hbar$-wave, see Fig.~\ref{fig:hbar}. Black circles mark the snapshot times.  
The solid black curve shows distinct echo peaks for the clean Dirac system, Eq.~\eqref{eq:Dirac}. The insets show sketches of the Dirac-type dispersion with a gap opening at $t=t_0$. 
Further curves correspond to different disorder types: gap disorder with $\tau_\text{gap} \approx 0.2 t_0 $
(red dotted), spatial disorder with $\tau_{\text{imp}} \approx 0.8 t_0$ (blue) and $\tau_{\text{imp}} \approx 0.2 t_0$ (green). 
$\tau_{\text{imp}}$ and $\tau_{\text{gap}}$ are the respective elastic scattering times (see Sec. \ref{app:disorder} in the Appendix).
(b) Modulus $|A(k)|$ of the transition amplitude, Eq.~\eqref{eq:GRA-transprob}, plotted as a function of parameters $\eta$ and $\mu$.} 
\label{fig:graphene}
\end{figure*}

The two-dimensional (2D) Dirac system considered above could describe fermions 
in real \cite{castroneto2009} and artificial \cite{tarruell2012} graphene, 
or e.~g.~Dirac plasmons in metallic nanoparticle lattices \cite{weick2013} and polaritons in a honeycomb lattice \cite{Jacqmin2014}.
In such systems the velocity is approximately constant -- it does not depend on the $k$-vector -- and equal in
magnitude, but opposite in direction, on the upper and lower Dirac cones.
Our $t$-inversion protocol aims at inducing a ``population reversal'', say from the upper to the lower Dirac cone,
which corresponds to an inversion of the velocity and thus effectively a propagation back in time.  
This is achieved by applying a short, spatially extended and (roughly) uniform perturbation
opening a gap in the spectrum: Once initial upper cone states suddenly find themselves in the forbidden gap region,
they start coherent oscillations between the upper and lower branches of the spectrum
-- akin to those responsible for Zitterbewegung.
A proper tuning of such oscillations ensures that, by the time the perturbation is switched off and the gap closes,
the states will end up in the lower Dirac cone.
\footnote{In fermionic systems such lower cone states must be available, i.e. they should not be already occupied, 
see discussion following Eq.~\eqref{eq:GRA-transprob}.}  
A protocol of this kind must ensure that the initial wave function,
a arbitrary coherent superposition of particle-like $k$-modes at different (positive) energies,
keeps its shape while reforming as a coherent superposition of hole-like $k$-modes in the corresponding (negative) energy window,
and that the probability for this ``rigid'' transition to the hole branch is as high as possible. 
As we will see, the linearity of the dispersion plays here the critical role.
While the physical gap-opening mechanism depends on the Dirac-like system
considered \footnote{For example, gaps in real graphene can be dynamically opened and closed via THz radiation 
\cite{fistul2007,scholz2013,usaj2014}.  A further experimental example of ultrafast manipulation of the graphene spectrum
appears in Ref.~\cite{higuchi2016}.}, its single crucial requirement is its non-adiabatic character as quantified below.

The effective Dirac Hamiltonian reads
\begin{equation}
 H= a \vk \cdot \vsigma + M(t) \sigma_z = H_0+H_1, 
\label{eq:Dirac}
\end{equation}
where the mass term
\begin{equation}
M(t) = \left\{\begin{array}{ll} M_0 \,, \quad & t_0<t<t_0+\Delta t, \\ 
	0 \,, & \text{otherwise},
\end{array}
\right.
\label{eq:M}
\end{equation}
acts as the time-dependent perturbation which temporarily opens a gap.  The eigenenergies and eigenstates of $H_0$ 
are $E_{k,\pm} = \pm a k$ and $\psi_\pm(\vk) = \frac{1}{\sqrt{2}}\left(\begin{array}{c} 1 \\ \pm e^{i\theta_{\vk}}\end{array}\right)$, 
with $\theta_{\vk}$ the polar angle in $k$-space. 
During the pulse the eigenenergies of $H$ are $\varepsilon_{k,\pm} = \pm\sqrt{M_0^2+E_{k,\pm}^2}$.
In the time interval $\Delta t$, an initial $H_0$ eigenstate is subject to oscillations, 
whose cycle depends on the pulse strength $M_0$ and duration $\Delta t$.   
The amplitude $A$ to end up in the counter-propagating $H_0$ eigenstate at $t= t_0+\Delta t$ after the pulse can be tuned
by adjusting both parameters.  A straightforward calculation yields (see Appendix \ref{app:derivation})
\begin{equation}
 \begin{split}
 A(k) &= \langle\psi_{\pm, \vk}\mid e^{-\frac{i}{\hbar} H \Delta t}\mid \psi_{\mp, \vk} \rangle  \\
      &= -\frac{i}{\sqrt{1+\eta^2}} \sin\left(\mu \sqrt{1+\eta^2}\right), \label{eq:GRA-transprob}
 \end{split}
\end{equation}
where we introduced the dimensionless parameters $\eta=E_{k,\pm}/M_0$ and $\mu = M_0 \Delta t/\hbar$.
The amplitude need be maximized for optimal echo strength, which then requires $\eta\ll1, \mu\approx\pi(n+1/2)$, $n\in\mathbb{Z}$,
corresponding to $E_{k,\pm}\ll M_0 \approx \pi\hbar/2\Delta t$ for $n=0$.
Notice that in a fermionic system the counter-propagating states need be empty.  This is ensured e.g. if Fermi level
(before injection of the wavepacket) is such that $\epsilon_F < -M_0$.
The reversal amplitude ~\eqref{eq:GRA-transprob}
is weakly $k$-dependent, a critical characteristic tied to the linearity of the dispersion,
\footnote{The QTM principle works also in non-linear dispersions, though less effectively.
A discussion of this physics will appear elsewhere.} which suggests the QTM to be highly effective in a wide $k$/energy range.  
As we will see shortly, the numerical simulations confirm this.  
In general, given an initial wave packet $\psi({\bf r},0)=(2\pi)^{-2}\int\,{\rm d}^2{\bf k}\,  \psi({\bf k},0) e^{i{\bf k}\cdot{\bf r}}$, 
a convenient measure of the echo strength is given by the correlation 
\begin{equation}
\mathcal{C}(t) = \int \,{\rm d}^2\br \, |\psi(\br,0)| |\psi(\br,t)| \label{eq:spaceCorr}
\end{equation}
between the moduli of the amplitudes at times $0$ and $t$.

To illustrate the QTM effect, a complicated wave packet resembling $\hbar$ is numerically propagated in time (see Fig.~\ref{fig:hbar}).
%by a wave packet propagation algorithm (see Methods).  
At times $t_0$ and $2.5 t_0$, a pulse with $M_0 = 8 \langle E_k \rangle$ and $\mu = M_0\Delta t/\hbar=\pi/2$ is applied, with $\Delta t\ll t_0$.
Here $\langle E_k \rangle = a \langle k \rangle$ is the mean wave packet energy, {\it i.e.}\ $\langle \eta\rangle = 1/8$. 
The snapshots in Fig.~\ref{fig:hbar} demonstrate that even after full destruction the spatial distribution of the initial wave packet can be reconstructed. 
This is quantified and confirmed in Fig.~\ref{fig:graphene}a), showing the corresponding correlator \eqref{eq:spaceCorr}. 
At the echo time ($t=2t_0+\Delta t \approx 2 t_0$), major parts of the time propagated wave packet indeed return to the initial position. 
This QTM mechanism is not limited to a single pulse: subsequent kicks at $t=2.5t_0$ and $t=3.25t_0$ 
cause further peaks, albeit of decreasing size.
The distinct echo peaks, based on the linear dispersion relation, arise since the kinetic phases accumulated by each 
$k$-mode during forward ($0\rightarrow t_0$) and backward ($t_0+\Delta t\rightarrow2t_0+\Delta t$) propagation add up to zero (see Appendix \ref{app:derivation}).

The numerical simulations are based on the wave packet propagation algorithm 
Time-dependent Quantum Transport (TQT) \cite{krueckl2013}. The state is discretized on a square grid and the time evolution is calculated for sufficiently small time steps such that the Hamilton operator can be assumed time independent for each step. We calculate the action of $H$ on $\psi$ in a mixed position and momentum-space representation by the application of Fourier Transforms. With this a Krylov Space is spanned, which can be used to calculate the time evolution using a Lanczos method \cite{lanczos1950}. 

For an arbitrary wave packet, the echo strength $\mathcal{C}(2t_0+\Delta t)$ is analytically given solely in terms of the amplitude $A(k)$, Eq.~\eqref{eq:GRA-transprob},
and the wave packet at $t\!=\!0$ as
\begin{equation}
\label{eq:GRA-transprob2}
\mathcal{C}(2t_0+\Delta t) =  \int \,{\rm d}^2\br \, \Big|\psi(\br,0)\Big| \, \Big|\int \frac{{\rm d}^2\vk}{(2\pi)^2}  \, A(k)\psi(\vk,0)e^{i\vk\cdot\br}\Big|.
\end{equation}

Figure \ref{fig:graphene}b) shows the $\eta$- and $\mu$-dependence of $|A(k)|$,
obtained from the time-reversal amplitude \eqref{eq:GRA-transprob}. One finds extended stripes of high fidelity.
To check this analytical result we simulate the propagation of a normalized 2D Gaussian wave packet with positive energy 
and small $k$-space width $\Delta k\ll k_0$ compared to the absolute value $k_0$ of its mean wave vector, 
such that $A(k)\approx A(k_0)$ for all $k$-modes involved.
Under these assumptions, Eq.~\eqref{eq:GRA-transprob2} reduces to $\mathcal{C}(2t_0+\Delta t) \approx|A(k_0)|$ (see Appendix \ref{app:derivation}),
%  in Supplementary Information~\ref{supp-app:derivation}),
which can be  compared with the correlation $\mathcal{C}(2t_0+\Delta t)$, Eq.~\eqref{eq:spaceCorr}, obtained from full numerical time evolution.  
As the mean difference between analytics and numerics is $0.03$, only the analytical plot is shown.
Clearly, strong echoes can be obtained in the full energy range $0\lesssim\eta\lesssim1$.

\section{Disorder}
\label{sec:disorder}

We now investigate the QTM robustness against disorder.
While this is typically present in a real system, it should be emphasized that state-of-the-art
hBN-encapsulated graphene samples are effectively ballistic over scales of several microns, \cite{wang2013}
corresponding to (transport) scattering times of several picoseconds.
For the sake of clarity, we keep the discussion at a qualitative level and refer the reader to Appendix \ref{app:disorder} for quantitative details.
We consider two types of disorder: a static spatial disorder potential and a spatially random pulse strength (referred to as ``gap disorder'').  
Spatial disorder enters into the Hamiltonian \eqref{eq:Dirac} as a time- and (pseudo)spin-independent potential 
$V_{\text{imp}}(\br)\sigma_0$, where $\sigma_0$ is the unit matrix in (pseudo)spin-space.  
Gap disorder is instead given by $V_{\text{gap}}(\br) \sigma_z$ for $t\in\lbrack t_0, t_0 +\Delta t\rbrack$, 
{\it i.e.}~only during the pulse.  Both random potentials are Gaussian distributed with width $u_{\text{imp}}$ or $u_{\text{gap}}$.  
As shown in Fig. \ref{fig:graphene}a), the echo is clearly more sensitive to a static random impurity potential than to gap disorder.
This is expected, and can be understood within the framework of Loschmidt echo theory \cite{jalabert2001,calvo2008,goussev2012b}: 
If a $t$-inversion protocol is not perfect, the echo signal decays as a function of the propagation time $t_0$.  Spatial disorder reduces the fidelity, 
since the QTM mechanism, even for an optimally calibrated pulse, achieves ``population reversal'' without directly affecting
the impurity scattering dynamics.  In other words, the QTM protocol does not lead to $V_{\text{imp}}(\br) \rightarrow -V_{\text{imp}}(\br)$.
In this sense elastic disorder has a qualitatively similar effect to inelastic scattering, whose effects are also
not undone by the QTM.  Gap disorder plays in principle a similar QTM-breaking role.  
However, and contrary to spatial disorder, it is active only during the very short 
pulse duration time $\Delta t \ll t_0$ and thus causes only negligible echo losses, reflected in the perfect agreement of the black line and dotted red line in Fig.~\ref{fig:graphene}a).
Our analysis also highlights a fundamental difference between standard spin echoes
and the present wave function echo: while a spin echo decays because of dynamical (inelastic) perturbations leading to $T_2$,
the role of $T_2$ is here played by the elastic scattering time $\tau$.  This suggests
an application of the QTM as a probe of the quality of a sample -- much as the spin echo is used as a probe
of decoherence in two-level systems.

We now explain at a qualitative level how and why static (elastic) disorder affects our echo,
whereas spin echoes are insensitive to it.
First, consider the scattering off a single impurity, assuming an incoming plane wave with a given propagation direction $\hat{\bf v}$.
Scattering leads to a position-dependent change of the wave front propagation direction. 
Considering true time reversal after the scattering process, every scattered part of the wave propagates back to the impurity and is scattered again.
However, due to destructive interference only the (inverted) initial propagation direction $-\hat{\bf v}$ survives.

In the presence of many scatterers, a Feynman path approach provides convenient insights.
While a phase $\varphi_s$ is accumulated along one particular path $s$ due to scattering off impurities, the same phase with inverted
sign $-\varphi_s$ is picked up on the way back, after (perfect) time inversion due to the time reversal
operator $\mathcal{T}\propto\mathcal{C}\sigma_y$, where $\mathcal{C}$ indicates complex conjugation.
Every backward path $s'$ other than the original one leads to a different phase $-\varphi_{s'}\neq-\varphi_s$. 
This causes destructive interference and ensures that only the contribution from the original path $s$ survives.

This phase inversion is also achieved in spin echoes.  Depending on the environment, the spins precess slower or faster.  
By applying a $\pi$-pulse at time $t_0$ which flips the spin, the faster spins are ``suddenly behind'' the slower ones.  
Neglecting inelastic effects, all spins are in phase again at $2t_0$ leading to the Hahn echo \cite{hahn1950}.

As opposed to the discussion about perfect time reversal, our pulse does not define an exact $t$-inversion protocol even in the absence of inelastic scattering, 
since it only inverts (``flips'') the kinetic phase due to $H_0$, which is $e^{-iE_\pm t_0 /\hbar}$.
Without disorder this is the only phase present, and thus it disappears for a closed loop (forward, then backward) propagation 
(see Appendix ~\ref{app:derivation}).  
With disorder the phase due to the random potential $V_{\text{imp}}(\vq)$ is not inverted and therefore not cancelled after the pulse on the way back.  
This leads to a ``dephasing'', such that contributions from various paths $s'$ survive at each impurity as opposed to the perfect time reversal discussed above, where only the time-reversed counterpart of the incoming path survives.

%%%%%%%%%%%%%%%%%%%%%%%%%%%%%%%%%%%%%%%%%%%%%%%%%%%%%%%%%%%%%%%%%%%%%%%%%%%%%%
% 
\section{Conclusions}
\label{sec_conclusion}

The analytical and numerical considerations presented in this work confirm the principles behind our QTM for pseudo-relativistic graphene-like systems.
This means that a sufficiently fast and spatially extended perturbation which opens a gap in a Dirac system
can act similar to a microwave $\pi$ pulse in spin-echo experiments, effectively $t$-inverting the orbital wave function dynamics
and thus generating a wave function echo.  
% Topological insulators \cite{hasan2010} open up the intriguing perspective 
% of unifying spin and orbital echoes: States at the surface of a 3D topological insulator are governed by an
% effective Dirac equation very similar to graphene where, most notably, the electron spin takes the role of graphene pseudospin. 
% We applied the QTM concept, presented here in detail for graphene, also to wave packet dynamics governed by the Dirac Hamiltonian
% \begin{equation}
%  H= a (\sigma_x k_y - \sigma_y k_x) + M(t) \sigma_z
% \label{eq:Dirac-TI}
% \end{equation}
% for topological insulator surfaces states \cite{hasan2010}.
% The results are practically identical to those for graphene, representing a simultaneous echo in the combined Hilbert space 
% of real spin and spatial degrees of freedom.
% \footnote{In topological insulators the gap $M(t)$, Eq.~(\ref{eq:M}), could in principle be opened simply via Zeeman coupling to a pulsed
% perpendicular magnetic field.  We are however not aware of currently available materials with a large enough
% out-of-plane $g$-factor.}

It is important to remark that the QTM does not require time-reversal symmetry to be preserved.
Indeed, we checked both analytically and numerically that high-fidelity echoes can also be obtained e.g. in graphene in the presence of a constant perpendicular magnetic field ${\bf B}=\nabla \times {\bf A}$,
described by the Hamiltonian
$H = a \left[\vk-(e/\hbar){\bf A}\right]\cdot \vsigma + M(t) \sigma_z.$
Wave packets in this case consist of superpositions of Landau levels with discrete energies $E_{n,\pm} \propto \pm\sqrt{n}$ \cite{clure1956, goerbig2011}, 
while during the pulse the dispersion becomes $\varepsilon_{n,\pm} = \sqrt{M_0^2+E_{n,\pm}^2}$.
The analogy with the discussion of Eq.~\eqref{eq:Dirac} is evident, and in fact
for each electron-like (upper Dirac branch) Landau level, the transition amplitude to its hole-like (lower Dirac branch) equivalent is again given by Eq.~\eqref{eq:GRA-transprob},
where now $\eta = E_{n,\pm} / M_0$. 
Since the propagation directions of electron- and hole-like Landau levels are reversed, the QTM principle still applies,
and strong echoes are obtained for $0\lesssim \eta \lesssim 1$.

The fact that time-reversal invariance need not be preserved may allow a further twist to the QTM proposal. 
The idea is to exploit the orbital effects of a pulsed out-of-plane magnetic field, rather than a mass gap pulse:
when the magnetic field is switched on the Dirac $k$-dispersion is abruptly changed to a gapped Landau level spectrum,
suggesting the feasibility of a Landau level-based QTM \cite{SuperPhillipp}.

Various experimental realizations of the QTM proposed here can be imagined.
Our QTM represents a general proof of concept, based on a single-particle picture and
including the assumption that the inelastic relaxation time of the injected wave packet is larger than $t_0$.
This condition imposes certain restrictions in real graphene \cite{urich2011,song2015,tielrooij2015},
though femtosecond laser pulses, routinely employed in nano-spectroscopy,\cite{eisele2014} could be
employed to open gaps therein, and indeed recent experimental advances suggest such restrictions not to be critical \cite{higuchi2016}.
On the other hand, certain forms of artificial graphene \cite{tarruell2012,Jacqmin2014} could be amenable 
to a straightforward experimental implementation of the QTM.
We also note that similar physics can be expected in the surface states of 3D topological insulators,\cite{hasan2010}
though this will be discussed elsewhere.

We furthermore demonstrated that the $t$-inversion protocol does not require time-reversal
symmetry and is practically insensitive to pulse (gap) disorder.
Vice versa, QTM-based echo spectroscopy could be used as a sensitive local probe of elastic and inelastic scattering times in Dirac-type systems.

In summary, we have proposed an instantaneous QTM for an extended quantum state,
based on the pseudo-relativistic dispersion of (bosonic or fermionic) Dirac-like systems.
An experimental realization of such an echo mechanism is within reach in state-of-the-art real or artificial graphene.

%%%%%%%%%%%%%%%%%%%%%%%%%%%%%%%%%%%%%%%%%%%%%%%%%%%%%%%%%%%%%%%%%%%%%%%%%%%%%%

\begin{acknowledgments}
We thank R. Huber for useful discussions.
A.G. acknowledges the support of EPSRC Grant No. EP/K024116/1. C.G., V.K., K.R. and P.R. acknowledge support from Deutsche Forschungsgemeinschaft within SFB 689 and GRK 1570.
\end{acknowledgments}

\appendix

\section{Derivation of transition probability to counter-propagating eigenstate}
\label{app:derivation}

We first derive Eq.~\eqref{eq:GRA-transprob} describing the transition amplitude owing to a time-dependent pulse in a Dirac-type system. The effective Hamiltonian for graphene-like (single-cone) systems is given by Eqs.~\eqref{eq:Dirac} ~and \eqref{eq:M}.
The eigenenergies and eigenstates of $H_0$ are 
\begin{align}
&E_\pm = \pm a k \\
&\psi_\pm(k) = \frac{1}{\sqrt{2}}\left(\begin{array}{c} 1 \\ \pm e^{-i\theta_k}\end{array}\right),
\end{align}
where $\theta_k$ is the polar angle in $k$-space. 
During the time interval $\Delta t$, an initial $H_0$ eigenstate is subject to Rabi-like oscillations, 
whose cycle depends on the pulse's strength $M_0$ and length $\Delta t$. The new eigenenergies and states become 
\begin{align}
&\varepsilon_\pm = \pm \sqrt{a^2 k^2+M_0^2}, \\
&\chi_\pm(k) =\frac{1}{\sqrt{a^2k^2+\left(M_0+\varepsilon_\pm\right)^2}}
\left(\begin{array}{c} M_0 + \varepsilon_\pm \\ ak \,e^{-i\theta_k}\end{array}\right).
\end{align}  
Thus, a band gap opens at $k=0$ with width $\Delta = 2M_0$.

The time evolution is explicitly performed for one mode, so as to derive the transition probability.  Consider, an initial eigenstate with negative energy $|\phi^k(t=0)\rangle=|\psi^k_-\rangle$. The index $k$ is from now on omitted
for sake of brevity.  The time evolution up to $t=t_0$ is trivial and results in a global phase: $|\phi(t)\rangle = e^{-\frac{i}{\hbar}E_- t_0}|\psi_-\rangle$.
During the pulse, the time evolution is governed by $H$, therefore we decompose $|\psi_-\rangle$ into eigenstates $|\chi_\pm\rangle$:
\begin{equation}
|\psi_-\rangle = \sum\limits_{s=\pm}\alpha_s |\chi_s\rangle 
\label{suppeq:initialstate}
\end{equation}
with 
\begin{equation}
\alpha_s = \langle \chi_s\mid\psi_-\rangle
\end{equation}
and the time evolution from $t=t_0$ to $t=t_0+\Delta t=t_1$ becomes
\begin{align}
|\phi(t_1)\rangle &= e^{-\frac{i}{\hbar}E_- t_0}e^{-\frac{i}{\hbar}H \Delta t}|\psi_-\rangle \nonumber \\ &= e^{-\frac{i}{\hbar}E_- t_0}\sum\limits_{s=\pm}\alpha_s e^{-\frac{i}{\hbar}\varepsilon_s \Delta t} |\chi_s\rangle. 
\end{align}
We are interested only in the component propagating back to its initial position, 
thus we project $ |\phi(t_1)\rangle$ onto $|\psi_+\rangle$, 
which has opposite velocity as compared to the initial state,
\begin{align}
\langle \psi_+ \mid \phi(t_1)\rangle &= e^{-\frac{i}{\hbar}E_- t_0}\sum\limits_{s=\pm}\alpha_s e^{-\frac{i}{\hbar}\varepsilon_s \Delta t} \langle \psi_+ \mid\chi_s\rangle \nonumber \\ &= e^{-\frac{i}{\hbar}E_- t_0}\sum\limits_{s=\pm}\alpha_s e^{-\frac{i}{\hbar}\varepsilon_s \Delta t} \beta_s^*
\end{align}
with 
\begin{equation}
\beta_s = \langle \chi_s\mid\psi_+\rangle.
\end{equation}
The component $\langle \psi_- \mid \phi(t_1)\rangle$ keeps propagating in its initial direction and is lost for the echo.

The echo takes place at $t=2t_0+\Delta t\simeq2t_0$ ($\Delta t \ll t_0$), because the absolute value of the velocity is the same for $|\psi_+\rangle$ and $|\psi_-\rangle$.  
The last propagation step to $t_2$ is again trivial and yields an additional phase $e^{-\frac{i}{\hbar}E_+ t_0}$ for the component traveling back, 
which cancels with the phase from $t=0$ to $t=t_0$ ($E_+=-E_-$).  The echo amplitude thus reads
\begin{equation}
\langle \psi_+ \mid \phi(t_2)\rangle = \sum\limits_{s=\pm}\alpha_s\beta_s^* e^{-\frac{i}{\hbar}\varepsilon_s \Delta t} .\label{eq:finalstate}
\end{equation}
Inserting the explicit expressions for $\alpha_s$, $\beta_s$ and $\varepsilon_s$, a straightforward but tedious calculation yields
\begin{eqnarray}
\langle \psi_+ \mid \phi(t_2)\rangle &=& -\frac{i}{\sqrt{1+\frac{a^2k^2}{M_0^2}}} \sin\left(\frac{M_0 \Delta t}{\hbar} \sqrt{1+\frac{a^2k^2}{M_0^2}}\right)
\nonumber\\
\label{echoamplitude1}
&=& - \frac{i}{\sqrt{1+\eta^2}} \sin\left(\mu \sqrt{1+\eta^2}\right), \label{eqsupp:GRA-transprob}
\end{eqnarray}
where the dimensionless parameters $\eta = ak/M_0$ and $\mu = M_0 \Delta t/\hbar$ were introduced.
Equation \eqref{echoamplitude1} corresponds to Eq.~\eqref{eq:GRA-transprob}.
The trivial time evolutions before and after the pulse cancel each other in the absence of disorder, 
so that the echo strength is solely due to the
(modulus of the) transition amplitude to the counter-propagating eigenstate. 

Starting in $| \psi_+\rangle$ instead of  $| \psi_-\rangle$, or in a superposition of the two, leads to the same conclusions.

For an arbitrary initial wave packet $\phi({\bf r},0)=(2\pi)^{-2} \int\,{\rm d}^2k \phi({\bf k},0) e^{i{\bf k}\cdot{\bf r}}$,
the simulation-derived echo strength $\mathcal{C}(2t_0)$ can be analytically estimated.  In equations~\eqref{suppeq:initialstate}-\eqref{eqsupp:GRA-transprob}, we solve the Schr\"odinger equation exactly for the population-reversed part of the wave packet, {\it i.e.}\ the only part contributing to the echo is
\begin{equation}
\phi(\vk,2t_0+\Delta t) = A(k) \phi(\vk,0),
\end{equation}
and therefore in real space 
\begin{align}
\phi(\vq, 2t_0+\Delta t) &=  \int\frac{{\rm d}^2k}{(2\pi)^2} \,\phi({\bf k},2t_0) e^{i{\bf k}\cdot{\bf r}} \nonumber \\ &= \int\frac{{\rm d}^2k}{(2\pi)^2}  \,A(k) \phi({\bf k},0) e^{i{\bf k}\cdot{\bf r}}.
\end{align}
As noted, the kinetic phases accumulated before and after the pulse cancel at $t=2t_0+\Delta t \simeq 2 t_0$, preventing interference effects in real space. Moreover, the linear band structure with constant phase velocity keeps the shape of the wave packet.
% All this knowledge leads analytically to the space allows us to  with the wave packet in $k$-space at time $2t_0$: 

Thus, our analytical estimate yields
\begin{align}
\mathcal{C}(2t_0) &= \int\,{\rm d}^2\vq \, |\phi(\vq,0)| |\phi(\vq,2t_0)| \nonumber \\ &=  \int\,{\rm d}^2\vq \, \Big|\phi(\vq,0)\Big| \,\Big|\int\frac{{\rm d}^2k}{(2\pi)^2}  \,A(k) \phi({\bf k},0) e^{i{\bf k}\cdot{\bf r}}\Big| \nonumber \\ &=: \mathcal{C}_A \,,
\end{align}
which is the same as Eq.~\eqref{eq:GRA-transprob2}.

In case of nearly constant $A(k)\approx A(k_0)$, e.g.\ for narrow peaked wave packets in $k$-space around an average value $k_0$, the echo strength can be estimated by
\begin{align}
\mathcal{C}(2t_0) &\approx \int \,{\rm d}^2\vq \, \Big|\phi(\vq,0)\Big|\, \Big|\int\frac{{\rm d}^2k}{(2\pi)^2}\,A(k_0) \phi({\bf k},0) e^{i{\bf k}\cdot{\bf r}}\Big| \nonumber \\ &=A(k_0)\int \,{\rm d}^2\vq \, |\phi(\vq,0)| \, | \phi({\bf r},0)| = A(k_0),
\end{align}
having taken a normalized initial wave packet.

\section{Disorder effects to the echo}
\label{app:disorder}

\subsection{Spatial disorder}
For the numerical investigation of disorder effects, we use a random, pseudospin-independent potential $V_{\text{imp}}(\vq)\sigma_0$. 
The latter assigns to every grid point $i$ a normally distributed value $\beta_i$, which is then multiplied with the disorder strength $u_{\text{imp}}$.  
The discontinuous potential is then smoothed by a Gaussian distribution with width $l_0$.  This leads to
\begin{equation}
V_{\text{imp}}(\vq) = \frac{u_{\text{imp}}}{\mathcal{N}} \sum\limits_i \beta_i e^{-\frac{(\vq-\vq_i)^2}{l_0^2}},
\end{equation}
where the sum runs over all grid points. The normalization $\mathcal{N}$ is due to numerical reasons and is given by 
\begin{equation}
\mathcal{N} = \left[\frac{1}{A}  \int_A \,{\rm d}^2r \left(\sum\limits_{i} \beta_i e^{-\frac{(\vq-\vq_i)^2}{l_0^2}} \right)^2 \right]^\frac{1}{2},
\end{equation}
$A$ being the (finite) grid area for the numerical simulation.
The correlator
\begin{equation}
\langle V_\mathrm{imp}(\vq)V_\mathrm{imp}(\vq^\prime)\rangle = u_{\text{imp}}^2 e^{-\frac{(\vq-\vq^\prime)^2}{2l_0^2}},
\end{equation}
where $\langle\cdot\rangle$ stands for disorder average, is needed in order to compute the scattering time.  The latter is given by 
(see e.g.~\cite{akkermansbook}) 
\begin{equation}
\frac{\hbar}{\tau_k} =  \int \frac{{\rm d}\vk^\prime d\vq}{2\pi\hbar} \delta(k-k^\prime) \langle V_\mathrm{imp}(0)V_\mathrm{imp}(\vq)\rangle e^{i \vq\cdot\vk^\prime },
\end{equation}
and can be calculated analytically as
\begin{equation}
\frac{1}{\tau_k} = \frac{2\pi}{a\hbar} u_{\text{imp}}^2 l_0^2 k\, e^{-l_0^2k^2} I_0(l_0^2k^2). \label{eq:scat_time}
\end{equation}
$I_0(x)$ is the modified Bessel function of zeroth kind.

\begin{figure*}[t]
\includegraphics[width=0.8\textwidth]{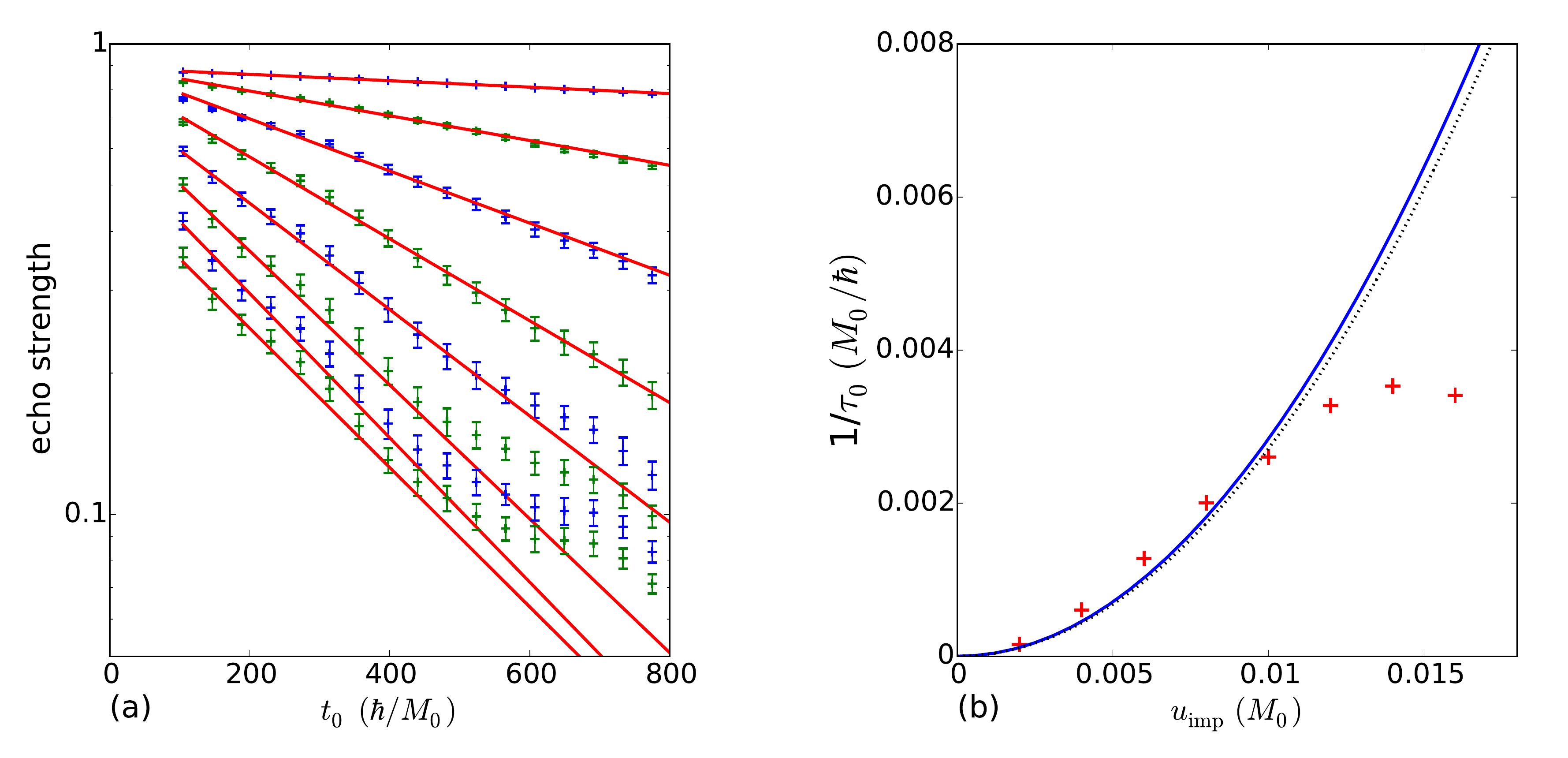}
\caption{Effects of disorder on graphene echo. (a) The echo strength measured by the echo fidelity $\tilde{M}(t_\text{echo})$ is shown as a function of pulse time $t_0$ for various disorder strengths $u_{\text{imp}}$ between $0.002M_0$ and $0.016M_0$, averaged over 50 realizations (see text). The error bars denote the standard error of the mean. An exponential fit is used to extract the decay rate. For higher $u_{\text{imp}}$ an expected saturation sets in, such that only a brief regime of exponential decay is visible.  (b) Plot of the fitted decay rates $1/\tau_{0}$ as a function of $u_{\text{imp}}$, Eq.~\eqref{eq:scat_time}, which is quadratic for weak scattering. The black dotted line is the analytically expected curve, which matches well the fitted quadratic function (blue), until the expected strong scattering saturation sets in.   \label{fig:loschmidt-echo}}
\end{figure*}

In the main part, we explained on a qualitative level the effect of disorder. Here, we discuss it quantitatively and for that reason we turn to the theory of Loschmidt echoes
(see e.g. \cite{goussev2012b,jacquod2001,jalabert2001,cerruti2002}), where the role of disorder
has been thoroughly studied and characterized.  In this context the echo is measured by the ``fidelity''
\begin{equation}
M(t) = |\langle\phi\mid e^{iH_a t/\hbar} e^{-iH_bt/\hbar} \mid \phi \rangle|^2,
\label{eq:fid}
\end{equation}
which is the overlap squared between the initial and final, {\it i.e.}\ time-evolved, state.
The time propagation is governed by $H_b$ until the time $t$, and by $-H_a$ thereafter, {\it i.e.}\ the time evolution is (``by hand'') explicitly inverted. (Note the difference to our protocol where the flow in time is not changed.)
Moreover, we use instead the correlation $\mathcal{C}(t)$, Eq.~\eqref{eq:spaceCorr}, to quantify the echo
in a physically transparent way -- as the overlap between initial and time-evolved local density.
In order to establish a connection with the Loschmidt echo theory, it is however more convenient to introduce the following
``echo fidelity'' $\tilde{M}$
\begin{equation}
\tilde{M}(t) = |\langle \sigma_z \phi \mid e^{-iHt/\hbar} \mid \phi \rangle |.
\label{eq:echofid}
\end{equation}
Notice the difference with $\mathcal{C}(t)$, where $|\phi|$ appears, and the $\sigma_z$ Pauli matrix: 
$\tilde{M}(t)$ is the overlap of the time propagated state with the initial one \textit{with flipped spinor}, 
since the returning part of the wave packet is in the flipped eigenstate of $H_0$.
In the golden rule decay regime \cite{goussev2012b}, the (mean) echo strength $\tilde{M}$ decays exponentially in time
\begin{equation}
\tilde{M}(t)  \sim e^{-\frac{t}{2\tau}} . %= e^{-\frac{t_0}{\tau}}.
\label{eq:echofid-decay}
\end{equation}
As mentioned, the pulse time-reverses the dynamics due to $H_0$ only, without affecting the dynamics arising from the impurity potential. 

This can be seen by splitting the time evolution operator in three parts: before the pulse, during the pulse and after the pulse. Furthermore, we consider only the part of the wave function which is reflected, which means that the time evolution during the pulse is given by the transition amplitude $A(k)$ times the operator $\sigma_z$, which maps any state to its energy-inversed counterpart.
\begin{align}
 \tilde{M}(2t_0) &= |\langle  \phi \mid  \sigma_z  e^{-\frac{i}{\hbar}  (H_0 + V_r) t_0} \sigma_z  A(k)   e^{-\frac{i}{\hbar} (H_0 + V_r) t_0}  \mid \phi \rangle |^2 \nonumber \\
&=|\langle  \phi \mid    e^{-\frac{i}{\hbar} \sigma_z  (H_0 + V_r) \sigma_z t_0}  A(k)   e^{-\frac{i}{\hbar}  (H_0 + V_r) t_0}  \mid \phi \rangle |^2
\end{align}
Since $\sigma_z \sigma_i \sigma_z = - \sigma_i$ for $i\in\{x,y\}$ and $H_0$ is a linear combination of $\sigma_x$ and $\sigma_y$, the sign in front of $H_0$ changes after the pulse, whereas the term related to the pseudospin-independent potential $V_r$ is not affected at all by the pulse.
\begin{align}
 \tilde{M}(2t_0)&=|\langle  \phi \mid    e^{-\frac{i}{\hbar}  (-H_0 + V_r)  t_0}  A(k)   e^{-\frac{i}{\hbar} (H_0 + V_r) t_0}  \mid \phi \rangle|^2 .%\nonumber\\
% &=|\langle  \phi \mid    e^{+\frac{i}{\hbar}  (H_0 - V_r)  t_0}  A(k)   e^{-\frac{i}{\hbar} (H_0 + V_r) t_0}  \mid \phi \rangle |^2 .
\end{align}
As there is a different sign in the propagation due to $H_0$ before and after the pulse, there is {\it effectively} a propagation
backwards in time. On the other hand, our $\sigma_z$-pulse cannot invert the time evolution due to the disorder potential, 
which causes disorder to ultimately destroy the echo.

Since the scattering occurs during the whole propagation time $2 t_0$, we expect the echo fidelity
\eqref{eq:echofid-decay} to decay as
\begin{equation}
\tilde{M}(2t_0)  \sim e^{-\frac{2 t_0}{2\tau}} = e^{-\frac{t_0}{\tau}}.
\end{equation}

This decay is confirmed in Fig.~\ref{fig:loschmidt-echo}a), where the echo fidelity (= echo strength), Eq.~\eqref{eq:echofid}, is shown as a function of the pulse time $t_0$.  The initial wave packet is a 2D-Gaussian with small $k$-space width $\sigma_k\ll k_0$ as compared 
to the mean wave vector $k_0$ , such that the $k$-dependence of the scattering time can be neglected ($\tau_k\approx\tau_{k_0}=:\tau_0$).  
The echo strength is calculated for 50 different realizations of the random disorder potential and averaged subsequently.
For large disorder strengths $u_{\text{imp}}$, a saturation regime is reached, in accordance with Loschmidt echo theory\cite{goussev2012b}. 
The decay rate is extracted by fitting an exponentially decaying function to the data, and compared in Fig.~\ref{fig:loschmidt-echo}b) to the analytically expected decay rate $1/\tau_{0}$ from Eq.~\eqref{eq:scat_time}, yielding a good agreement.  

The time-decay of $\mathcal{C}(t)$ is qualitatively similar though slower, because the phase differences between the initial and the propagated wave packet are neglected in the modulus, preventing eventual destructive interference.

\subsection{Gap disorder}

The gap disorder potential $V_{\text{gap}}(\vq) \sigma_z$ models fluctuations in the pulse strength 
and is therefore only active in the short time window $\Delta t$.
Figure~\ref{fig:graphene} a) shows that gap disorder has practically no effect on the echo as compared to spatial disorder.  
This is expected, as spatial disorder is active during a time $2t_0 \gg \Delta t$.  
Assuming similar scattering times for $u_{\text{imp}} = u_{\text{gap}}$, practically no gap disorder-induced scattering takes place
during the pulse, since $\tau_{0} \gg \Delta t$.
Moreover, spatial and gap disorder acts slightly differently.  The impurities lead to a randomization of the propagation direction, 
such that a smaller amount of the wave packet goes back to the initial position.  Gap disorder instead modulates in space the transition
probability to the counter-propagating eigenstate, but the propagation direction is not randomized.

\bibliography{biblio_QTM_PRB_final}

%\begin{thebibliography}{10} 
%\end{thebibliography}

\end{document}